\documentclass{article}

\PassOptionsToPackage{numbers, sort&compress}{natbib}

\usepackage[final]{neurips_2021}

\usepackage[utf8]{inputenc} 
\usepackage[T1]{fontenc}    
\usepackage{hyperref}       
\usepackage{url}            
\usepackage{booktabs}       
\usepackage{amsfonts}       
\usepackage{nicefrac}       
\usepackage{microtype}      
\usepackage{xcolor}         
\usepackage{graphicx}
\usepackage{subcaption}     
\usepackage{tikz}           
\usepackage{pgfplots}       
\usepackage{pgf}            
\usepackage{siunitx}        
\usepackage{enumitem}       
\usepackage{amsthm}         
\usepackage{wrapfig}        
\usepackage{arydshln}       
\usepackage{resizegather}
\usepackage{chngcntr}       
\usepackage{amssymb}        
\usepackage{placeins}       
\usepackage{etoolbox}       

\usepackage{amsmath,amsfonts,bm}
\usepackage{accents} 

\def\eqref#1{equation~\ref{#1}}

\def\1{\bm{1}}

\def\ve{{\bm{e}}}

\def\vh{{\bm{h}}}

\def\vm{{\bm{m}}}
\def\vn{{\bm{n}}}

\def\vr{{\bm{r}}}

\def\vx{{\bm{x}}}

\def\vz{{\bm{z}}}

\def\mF{{\bm{F}}}

\def\mH{{\bm{H}}}

\def\mR{{\bm{R}}}

\def\mW{{\bm{W}}}
\def\mX{{\bm{X}}}

\DeclareMathAlphabet{\mathsfit}{\encodingdefault}{\sfdefault}{m}{sl}
\SetMathAlphabet{\mathsfit}{bold}{\encodingdefault}{\sfdefault}{bx}{n}
\newcommand{\tens}[1]{\bm{\mathsfit{#1}}}

\def\tW{{\tens{W}}}

\def\gF{{\mathcal{F}}}
\def\gG{{\mathcal{G}}}

\def\gN{{\mathcal{N}}}
\def\gO{{\mathcal{O}}}

\def\gR{{\mathcal{R}}}

\def\sN{{\mathbb{N}}}

\def\sT{{\mathbb{T}}}

\newcommand{\E}{\mathbb{E}}

\newcommand{\R}{\mathbb{R}}

\newcommand{\Var}{\mathrm{Var}}

\newcommand{\Cov}{\mathrm{Cov}}

\DeclareMathOperator{\SO3}{SO(3)}
\DeclareMathOperator{\SOg}{SO}
\DeclareMathOperator{\Og}{O}

\newlength{\dtildeheight}

\newlength{\dcheckheight}

\usepackage[capitalize,nameinlink]{cleveref} 

\robustify\bfseries  
\newcommand\mcc[1]{\multicolumn{1}{c}{#1}}  
\usetikzlibrary{calc,shapes,arrows.meta,angles,quotes,bending}
\usepgfplotslibrary{fillbetween} 

\crefname{section}{Sec.}{Sec.}
\crefname{appendix}{App.}{App.}
\crefname{definition}{Def.}{Defs.}
\crefname{proposition}{Prop.}{Props.}

\newtheorem{theorem}{Theorem}
\newtheorem{proposition}{Proposition}

\newtheorem{lemma}{Lemma}

\pgfdeclarelayer{background}
\pgfdeclarelayer{foreground}
\pgfsetlayers{background,main,foreground}

\definecolor{uglyblue}{named}{blue}
\definecolor{uglygreen}{named}{green}
\definecolor{uglyred}{named}{red}
\definecolor{uglyyellow}{named}{yellow}

\definecolor{brewerpurple}{RGB}{117,112,179}
\definecolor{brewergreen}{RGB}{27,158,119}
\definecolor{brewerred}{RGB}{217,95,2}

\definecolor{blue}{RGB}{1, 115, 178}
\definecolor{orange}{RGB}{222, 143, 5}
\definecolor{green}{RGB}{2, 158, 115}
\definecolor{red}{RGB}{213, 94, 0}
\definecolor{pink}{RGB}{204, 120, 188}
\definecolor{yellow}{RGB}{236, 225, 51}
\definecolor{brown}{RGB}{202, 145, 97}
\definecolor{grey}{RGB}{148, 148, 148}
\definecolor{brightblue}{RGB}{86, 180, 233}

\def\sharecolor{red!20}
\def\quadcolor{brewergreen!100}
\def\tripcolor{brewerred!100}
\def\blockcoloremb{green!20}
\def\blockcolorres{pink!30}
\def\blockcolormsgpas{grey!30}
\def\blockcolorint{yellow!30}
\def\blockcoloratom{blue!40}
\def\blockcolorupdate{orange!30}
\def\atominteractioncolor{brightblue!20}

\title{GemNet: Universal Directional Graph Neural Networks for Molecules}

\author{%
  Johannes Gasteiger, Florian Becker, Stephan Günnemann \\
  Technical University of Munich, Germany\\
  \texttt{\{j.gasteiger,beckerf,guennemann\}@in.tum.de} \\
}

\begin{document}

\maketitle

\begin{abstract}
    Effectively predicting molecular interactions has the potential to accelerate molecular dynamics by multiple orders of magnitude and thus revolutionize chemical simulations. Graph neural networks (GNNs) have recently shown great successes for this task, overtaking classical methods based on fixed molecular kernels. However, they still appear very limited from a theoretical perspective, since regular GNNs cannot distinguish certain types of graphs. In this work we close this gap between theory and practice. We show that GNNs with spherical representations are indeed universal approximators for predictions that are invariant to translation, and equivariant to permutation and rotation. We then discretize such GNNs via directed edge embeddings and two-hop message passing, and incorporate multiple structural improvements to arrive at the geometric message passing neural network (GemNet). We demonstrate the benefits of the proposed changes in multiple ablation studies. GemNet outperforms previous models on the COLL, MD17, and OC20 datasets by \SI{34}{\percent}, \SI{41}{\percent}, and \SI{20}{\percent}, respectively, and performs especially well on the most challenging molecules. Our implementation is available online. \footnote{\url{https://www.daml.in.tum.de/gemnet}}
\end{abstract}

\section{Introduction}

Graph neural networks (GNNs) have shown great promise for predicting the energy and other quantum mechanical properties of molecules. They can predict these properties orders of magnitudes faster than methods from quantum chemistry -- at comparable accuracy. GNNs can thus enable the accurate simulation of systems that are orders of magnitude larger. However, they still exhibit severe theoretical and practical limitations. Regular GNNs are only as powerful as the 1-Weisfeiler Lehman test of isomorphism and thus cannot distinguish between certain molecules \citep{morris_weisfeiler_2019,xu_how_2019}. Moreover, they require a large number of training samples to achieve good accuracy.

In this work we first resolve the questionable expressiveness of GNNs by proving sufficient conditions for universality in the case of invariance to translations and rotations and equivariance to permutations; and then extending this result to rotationally equivariant predictions. Simply using the full geometric information (e.g.\ all pairwise atomic distances) in a layer does not ensure universal approximation. For example, if our model uses a rotationally invariant layer we lose the relative information between components. Such a model thus cannot distinguish between features that are rotated differently. This issue is commonly known as the ``Picasso problem'': An image model with rotationally invariant layers cannot detect whether a person's eyes are rotated correctly. Instead, we need a model that preserves relative rotational information and is only invariant to \emph{global} rotations. To prove universality in the rotationally invariant case we extend a recent universality result based on point cloud models that use representations of the rotation group $\SO3$ \citep{dym_universality_2021}. We prove that spherical representations are actually sufficient; full $\SO3$ representations are not necessary. We then generalize this to rotationally equivariant predictions by leveraging a recent result on extending invariant to equivariant predictions \citep{villar_scalars_2021}. We then discretize spherical representations by selecting points on the sphere based on the directions to neighboring atoms. We can connect this model to GNNs by interpreting these directions as directed edge embeddings. For example, the embedding direction of atom $a$ would be defined by atom $c$, resulting in the edge embedding $e_{ca}$. Updating the spherical representation of atom $a$ based on atom $b$ then corresponds to two-hop message passing between the edges $e_{ca}$ and $e_{db}$ via $e_{ba}$, with atoms $c$ and $d$ defining the embedding directions. This message passing formalism naturally allows us to obtain the molecule's full geometrical information (distances, angles, and dihedral angles), and the direct correspondence shows the model's expressiveness.

We call this edge-based two-hop message passing scheme \emph{geometric message passing}, and propose multiple structural enhancements to improve the practical performance of this formalism. Based on these changes we develop the highly accurate and sample-efficient geometric message passing neural network (GemNet). We furthermore show that stabilizing the variance of GemNet's activations with predetermined scaling factors yields significant improvements over regular normalization layers.

We investigate the proposed improvements in a range of ablation studies, and show that each of them significantly reduces the model error. These changes introduce little to no computational overhead over two-hop message passing. Altogether, GemNet outperforms previous models for force predictions on COLL by \SI{34}{\percent}, on MD17 by \SI{41}{\percent}, and on OC20 by \SI{20}{\percent} on average. We observe the largest improvements for the most challenging molecules, which exhibit dynamic, non-planar geometries.
In summary, our contributions are:
\setlist{nolistsep}
\begin{itemize}[leftmargin=*,itemsep=2pt]
    \item Showing the \textbf{universality} of spherical representations for rotationally equivariant predictions, and discretizing this theoretical model to two-hop message passing with directed edge embeddings.
    \item \textbf{Geometric message passing}: Symmetric message passing enhanced by geometric information.
    \item Incorporating all proposed improvements in the \textbf{Geometric Message Passing Neural Network (GemNet)}, which significantly outperforms previous methods for molecular dynamics prediction.
\end{itemize}

\section{Related work}

\textbf{Machine learning potentials.} Research on predicting a molecule's energy and forces (so-called machine learning potentials) started with hand-fitted analytical functions and then gradually moved towards fully learned models. Arguably, classical force fields are their very first instances. They use analytical functions with coefficients that were hand-tuned based on experimental data. A popular example for these is the Merck Molecular Force Field (MMFF94) \citep{halgren_merck_1996}. The next wave of methods used kernel ridge regression based on fixed, hand-crafted molecular representations \citep{bartok_gaussian_2010,chmiela_machine_2017,faber_alchemical_2018}. Finally, modern research mostly focusses on fully end-to-end learnable models based on GNNs \citep{gilmer_neural_2017,schutt_schnet:_2017}. These models can also be combined with molecular features from quantum mechanical calculations to improve performance \citep{qiao_orbnet_2020}. We consider this combination as orthogonal research.

\textbf{Directional GNNs.} We can also achieve equivariance and invariance to rotations without relying on group representations. Directional GNNs achieve this by representing directional information explicitly \citep{schutt_equivariant_2021} or in the form of angles \citep{gasteiger_directional_2020} and dihedral angles \citep{flam-shepherd_neural_2021,liu_spherical_2021}. Our work is focused on this class of models, showing their expressiveness and proposing an improved variant, GemNet.

\textbf{Expressiveness of GNNs.} A large part of GNN research has been focused on their (limited) expressiveness. \citet{xu_how_2019,morris_weisfeiler_2019} first proved that they are only as expressive as the Weisfeiler-Lehman test of isomorphism and \citet{garg_generalization_2020} showed the limitations of basic directional message passing. \citet{maron_universality_2019,morris_weisfeiler_2019,morris_weisfeiler_2020,kondor_covariant_2019} then investigated higher-order representations to circumvent this issue. Finally, \citet{maron_provably_2019,azizian_expressive_2020} showed that so-called folklore GNNs are the most expressive GNNs for a given tensor order.

\textbf{Equivariant neural networks.} Equivariance and invariance have recently emerged as one of the foundational principles of modern neural networks \citep{cohen_group_2016,cohen_general_2019}. This is especially relevant for models in physics, for which we often know the symmetries a priori. Equivariant models for the $\SO3$ group were first investigated in the context of spherical convolutions by \citet{cohen_spherical_2018,esteves_learning_2018,kondor_clebsch-gordan_2018}. These methods leverage group representations to achieve full equivariance. They were then transferred to the context of 3D point clouds and molecules by \citet{weiler_3d_2018,thomas_tensor_2018,anderson_cormorant:_2019}, and further developed by \citet{fuchs_se3-transformers_2020,batzner_se3-equivariant_2021,finzi_generalizing_2020}. Importantly, \citet{yarotsky_universal_2021} proved the universality of 2D convolutional networks, and \citet{bogatskiy_lorentz_2020} extended this result to general groups. \citet{maron_learning_2020} proved universality for models invariant to $S_n$ and equivariant to an additional symmetry. \citet{dym_universality_2021} combined these results to prove universality for the joined group of translations, rotations, and permutations. Apart from reflections this is the exact group relevant for general molecules.

\section{Universality of spherical representations}

GNNs for molecules typically incorporate directional information in one of two ways: Via $\SO3$ representations \citep{thomas_tensor_2018,anderson_cormorant:_2019} or by using directions in real space \citep{gasteiger_directional_2020,schutt_equivariant_2021}. Directions in real space are associated with the three-dimensional $S^2$ sphere, while the $\SO3$ group is double covered by the four-dimensional $S^3$ sphere. Directional representations thus use one degree of freedom less than $\SO3$ representations, making them significantly cheaper. And, as we will prove in this section, directional representations actually provide the same expressivity as $\SO3$ representations for predictions in $\R^3$. We achieve this by showing that the $\SO3$-based tensor field network (TFN) \citep{thomas_tensor_2018} variant used by \citet{dym_universality_2021} is equivalent to a similar model based on spherical representations, in the case of rotationally invariant predictions. We then generalize a recent result by \citet{villar_scalars_2021}, which lets us extend our theorem to the rotationally equivariant case. Afterwards, we relate this universality to directional GNNs by interpreting them as a discretization of spherical representations.

\textbf{Preliminaries.} We consider a point cloud with $n$ points (atoms), each associated with a position and a set of rotationally invariant features (e.g.\ atom types), defined as $\mX \in \R^{3 \times n}$ and $\mH_{\text{in}} \in \R^{h \times n}$. In this section we define model classes by sets of functions $\gF$. As a first step, we are interested in proving that the set $\gF$ defining our model is equal to the full set of functions $\gG'$ that are invariant to the group of translations $\sT^3$ and rotations $\SO3$, and equivariant to the group of permutations $S_n$. We denote the codomain of functions in $\gG'$ as $W_{\text{T}}^n$, where $W_{\text{T}}$ is some representation of $\SO3$. We denote a vector's norm by $x = \| \vx \|_2$, its direction by $\hat{\vx} = \vx / x$, and the relative position by $\vx_{ba} = \vx_b - \vx_a$. Proofs are deferred to the appendix. Note that this section is not intended as an introduction to the $\SO3$ group. For a concise introduction in the context of machine learning see e.g.\ \citet[Section 3]{weiler_3d_2018} or \citet{kondor_clebsch-gordan_2018}.

\textbf{Tensor field network.} In order to show the equivalence of the TFN to spherical representations, we first need to define this model. Following \citet{dym_universality_2021}, we split the model into two parts: Embedding functions $\gF_{\text{feat}}$ that lifts the input into an equivariant representation, and pooling functions $\gF_{\text{pool}}$ that aggregate the results of multiple embedding functions on each point and computes the model output. The overall model is then defined as the set of functions
\begin{equation}
    \gF_{K(D), D}^{\text{TFN}} = \{ f \mid f(\mX, \mH_{\text{in}}) = \sum_{k=1}^K f_{\text{pool}}^{(k)*}(f_{\text{feat}}^{(k)}(\mX, \mH_{\text{in}})), f_{\text{pool}}^{(k)} \in \gF_{\text{pool}}^{\text{TFN}}(D), f_{\text{feat}}^{(k)} \in \gF_{\text{feat}}^{\text{TFN}}(D) \},
\end{equation}
where $D \in \sN$ denotes the function's maximum polynomial degree, $K(D) \in \sN$ is chosen such that \cref{th:univ_tfn} is fulfilled (\citet{dym_universality_2021} only prove the existence of this function), and $f^*$ denotes elementwise application of $f$ on all points. We then define the set $\gF_{\text{pool}}^{\text{TFN}}$ as all rotationally equivariant linear functions on the $\SO3$ group, i.e.\ all $\SO3$ convolutions \citep{kostelec_ffts_2008}. Note that these are more expressive than the self-interaction layers used originally \citep{thomas_tensor_2018}.
The embedding functions $\gF_{\text{feat}}^{\text{TFN}}(D) = \{ \pi_2 \circ f^{(2D)} \circ \dots \circ f^{(1)} \mid f^{(i)} \in \gF_{\text{prod}}^{\text{TFN}} \}$ consist of an auxiliary function $\pi_2(\mX, \mH) = \mH$ and a series of tensor product functions (called convolution by \citet{dym_universality_2021}) $\gF_{\text{prod}}^{\text{TFN}} = \{ f \mid f(\mX, \mH) = (\mX, \tilde{\mH}^{\text{TFN}}(\mX, \mH)) \}$. The intermediate representations are $\mH \in W_{\text{feat}}^n$, where $W_{\text{feat}}$ is a representation of $\SO3$ indexed by the degree $l$ and the order $m$. For $\mH_{\text{in}}$ we have $l\!=\!m\!=\!0$. The main update is defined by
\begin{equation}
    \tilde{\mH}^{\text{TFN}(l_o)}_{am_o}(\mX, \mH) = \theta \mH^{(l_o)}_{am_o} + \sum_{l_f, m_f} \sum_{l_i, m_i} C_{(l_f, m_f), (l_i, m_i)}^{(l_o, m_o)} \sum_{b \in \gN_a} F^{(l_f)}_{\text{TFN},m_f}(\vx_b - \vx_a) \mH^{(l_i)}_{bm_i},
    \label{eq:tfn}
\end{equation}
where $\theta$ is a (learned) scalar and $\gN_a$ are the neighbors of point $a$. The Clebsch-Gordan coefficients $C_{(l_f, m_f), (l_i, m_i)}^{(l_o, m_o)}$ arise from decomposing the tensor product of two input $\SO3$ representations (the filter and input representations) into a sum of output representations. Their exact values are not relevant for this discussion. We index the output with degree $l_o$ and order $m_o$, the learned filter with $l_f$ and $m_f$, and the input with $l_i$ and $m_i$. $F^{(l)}_{\text{TFN},m}(\vx) = R^{(l)}(x) Y_{lm}(\hat{\vx})$ is a rotationally equivariant filter, with a (learned) radial part $R$, which is any polynomial of degree $\le D$, and the real spherical harmonics $Y_{lm}$ with degree $l$ and order $m$. The spherical harmonics are the basis for the Fourier transformation of functions on the sphere, analogously to sine waves for functions on $\R$. We can prove universality for TFNs by using the universality of polynomial regression and showing that TFNs can fit any polynomial (see \citet{dym_universality_2021} for details), resulting in:
\begin{theorem}[\citet{dym_universality_2021}]
    Consider the set of functions $\gG$ mapping $\R^{3 \times n + h \times n} \to W_{\text{\upshape T}}^n$ that are equivariant to rotations and permutations and invariant to translations. For all $n \in \sN$,
    \setlist{nolistsep}
    \begin{enumerate}[leftmargin=*,itemsep=2pt]
        \item For $D \in \sN_0$, every polynomial $p \in \gG$ of degree $D$ is in $\gF_{K(D), D}^{\text{\upshape TFN}}$.
        \item Every continuous function $f \in \gG$ can be approximated uniformly on compact sets by functions in $\bigcup_{D \in \sN_0} \gF_{K(D), D}^{\text{\upshape TFN}}$.
    \end{enumerate}
    \label{th:univ_tfn}
\end{theorem}

\textbf{Spherical networks.} Instead of intermediate $\SO3$ representations we now switch to spherical representations, which are functions on the sphere $\mH: S^2 \to \R$. We define the set of functions $\gF_{K(D), D}^{\text{sphere}}$ analogously to $\gF_{K(D), D}^{\text{TFN}}$. However, for $\gF_{\text{feat}}^{\text{sphere}}(D)$ we use
\begin{equation}
    \tilde{\mH}^{\text{sphere}}_a(\mX, \mH)(\hat{\vr}) = \theta \mH_a(\hat{\vr}) + \sum_{b \in \gN_a} F_{\text{sphere}}(\vx_b - \vx_a, \hat{\vr}) \mH_b(\hat{\vr}),
\label{eq:sphere}
\end{equation}
with the filter function $F_{\text{sphere}}(\vx, \hat{\vr}) = \sum_{l, m} R^{(l)}(x) \Re[Y_m^{(l)*}(\hat{\vx}) Y_{m}^{(l)}(\hat{\vr})]$, using the real part $\Re$ of the complex spherical harmonics $Y_m^{(l)}$. The set of pooling functions for invariant predictions is
\begin{equation}
    \gF_{\text{pool}}^{\text{sphere}} = \{ f \mid f(\mH) = \theta_{\text{pool}} \int_{S^2} \mH(\hat{\vr}) \,\text{d}\hat{\vr} \},
\end{equation}
with the learnable parameter $\theta_{\text{pool}}$.
We obtain the universality theorem by showing the equivalence between this model and TFN for rotationally invariant functions. The proof is based on the connection between spherical harmonics and the Clebsch-Gordan coefficients \citep[3.7.72]{sakurai_modern_1993} (see \cref{app:univ_sphere}).
\begin{theorem}
    Consider the set of functions $\gG'$ mapping $\R^{3 \times n + h \times n} \to W_{\text{\upshape T}}^n$ that are equivariant to permutations and invariant to translations and rotations. For all $n \in \sN$,
    \setlist{nolistsep}
    \begin{enumerate}[leftmargin=*,itemsep=2pt]
        \item For $D \in \sN_0$, every polynomial $p \in \gG'$ of degree $D$ is in $\gF_{K(D), D}^{\text{\upshape sphere}}$.
        \item Every continuous function $f \in \gG'$ can be approximated uniformly on compact sets by functions in $\bigcup_{D \in \sN_0} \gF_{K(D), D}^{\text{\upshape sphere}}$.
    \end{enumerate}
    \label{th:univ_sphere}
\end{theorem}

Next, we extend \cref{th:univ_sphere} to rotationally \emph{equivariant} functions. We do this by generalizing a recent result by \citet{villar_scalars_2021} to obtain (see \cref{app:univ_vec}):

\begin{theorem}
    Let $h \colon \R^{d \times n + h \times n} \to \R^{d \times n}$ be any function that is equivariant to permutations and rotations and invariant to translations. For all $a \in [1, n]$, let the set of relative vectors $\{\vx_{ca} \mid c \in [1, n]\}$ not span a $(d-1)$-dimensional space. Then there are $n - 1$ functions $f^{(c)} \colon \R^{d \times n + h \times n} \to \R^{n}$ such that
    \begin{equation}
        \vh_a(\mX, \mH) = \sum_{\substack{c=1\\c \neq a}}^n f_a^{(c)}(\mX, \mH) \vx_{ca},
    \label{eq:vec_pred}
    \end{equation}
    where $f^{(c)}$ is equivariant to permutations, but invariant to rotations and translations.
    \label{th:univ_vec}
\end{theorem}

This theorem lets us extend a rotationally invariant model to an equivariant one, while providing universality guarantees. Together, \cref{th:univ_sphere} and \cref{th:univ_vec} (with $d = 3$) thus show that we can approximate any rotationally equivariant function using only representations on the $S^2$ sphere. We thus do not need $\SO3$ representations, spin-weighted spherical harmonics \citep{esteves_spin-weighted_2020}, triplet embeddings, or complex-valued functions. This result puts theory back in line with practice, where the best results are currently achieved without relying on these more expensive representations \citep{schutt_equivariant_2021}.

\section{From spherical representations to directional message passing}

\textbf{Directional representations.} To use spherical representations in a model we first need to find a tractable description. Instead of using spherical harmonics, we propose to sample the representations in specific directions $\hat{\vr}_i$. If we look at recent models, we see that they implicitly use the directions to each atom's neighbors for this purpose, i.e.\ they embed the edges in the molecule's graph. These directions define an \emph{equivariant} mesh that circumvents the aliasing effects that would arise from fixed grids \citep{kondor_clebsch-gordan_2018}. \citet{schutt_equivariant_2021} flexibly define the directional mesh in each layer by aggregating directions, while \citet{gasteiger_directional_2020} and others use a fixed mesh for each atom. We can refine this mesh of directions e.g.\ by using more neighbors or by interpolating between directions. The approximation error of this directional mesh is related to the spherical harmonic expansion via the mesh norm and the separating distance between directions \citep{jetter_error_1999,keiner_efficient_2007}. Note that depending on the discretization scheme the resulting mesh might not provide a universal approximation guarantee.

\cref{eq:sphere} only defines the relationship for a fixed direction, while models commonly use different directional meshes for the input and output. We model the relationship between different directions using a convolution with a learned filter $F_2$, which can only improve expressiveness. Note that the input function is only defined for specific directions $\hat{\vr}_i$. To simplify the resulting equations we express this using Dirac deltas. Since the input and output are spherical functions, the used filter $F_2$ has to be \emph{zonal}, i.e.\ it has to be isotropic and depend on only one angle \citep{esteves_learning_2018}. This can be expressed as \citep{driscoll_computing_1994}
\begin{equation}
\begin{aligned}
    \tilde{\mH}^{\text{dir}}_a(\mX, \mH)(\hat{\vr}_o) &= \theta \mH_a(\hat{\vr}_o) + \int_{\SO3} \sum_{b \in \gN_a} F_{\text{sphere}}(\vx_{ba}, \mR \hat{\vn}) \sum_{i \in \gR_b} \mH_{bi} \delta(\mR \hat{\vn} - \hat{\vr}_i) F_2(\mR^{-1} \hat{\vr}_o) \,\text{d}\mR\\
    &= \theta \mH_a(\hat{\vr}_o) + \sum_{b \in \gN_a} \sum_{i \in \gR_b} F_{\text{sphere}}(\vx_{ba}, \hat{\vr}_i) \mH_{bi} F_2(\measuredangle \hat{\vr}_o \hat{\vr}_i),
\end{aligned}
\end{equation}
where $\gR_b$ denotes the directional mesh of atom $b$ with mesh directions denoted by $\hat{\vr}_i$, and $\hat{\vr}_o$ specifies the output direction. The integral vanishes due to the Dirac delta $\delta$.

\textbf{General filters.} To see the relationship to GNNs we furthermore need to generalize the filter $F_{\text{sphere}}(\vx_{ba}, \hat{\vr}_i)$. This filter only depends on the angle $\measuredangle \hat{\vr}_i \hat{\vx}_{ba}$ since it is rotationally invariant:
\begin{lemma}
    $F_{\text{\upshape sphere}}(\mR \vx, \mR \hat{\vr}) = F_{\text{\upshape sphere}}(\vx, \hat{\vr})$ for any rotation matrix $\mR$.
    \label{lem:filter_inv}
\end{lemma}
We can therefore substitute $F_{\text{sphere}}$ with a general learnable filter $F_1$ that is parametrized by this relative angle. Since $F_{\text{sphere}}$ arises as a special case we do not lose expressivity. We thus obtain
\begin{equation}
    \tilde{\mH}^{\text{gem}}_a(\mX, \mH)(\hat{\vr}_o) = \theta \mH_a(\hat{\vr}_o) + \sum_{b \in \gN_a} \sum_{i \in \gR_b} F_1(x_{ba}, \measuredangle \hat{\vr}_i \hat{\vx}_{ba}) F_2(\measuredangle \hat{\vr}_o \hat{\vr}_i) \mH_{bi}.
    \label{eq:gem}
\end{equation}
We have now arrived at a message passing scheme that is directly derived from a theoretical model with universal approximation guarantees but only requires relative directional information. To see the connection to GNNs we interpret these discretized spherical representations as edge embeddings pointing towards $\hat{\vr}_o$ and $\hat{\vr}_i$. \cref{eq:gem} then corresponds to two-hop message passing between the edge embeddings of $\hat{\vr}_o$ and $\hat{\vr}_i$ via the edge $\hat{\vx}_{ba}$. Interestingly, the central learnable part of \cref{eq:gem} is the product of the filters $F_1(x_{ba}, \measuredangle \hat{\vr}_i \hat{\vx}_{ba})$ and $F_2(\measuredangle \hat{\vr}_o \hat{\vr}_i)$ with the input representation, which is strikingly similar to the Hadamard product used in modern GNNs \citep{schutt_schnet:_2017,gasteiger_fast_2020} -- except that these only use one-hop message passing.

\section{Geometric message passing} \label{sec:gmp}

\begin{wrapfigure}[15]{r}{0.31\textwidth}
    \centering
    \vspace*{-0.3cm}

\begingroup
\medmuskip=0mu
\pgfdeclarelayer{background}
\pgfdeclarelayer{foreground}
\pgfsetlayers{background,main,foreground}
\begin{tikzpicture}
    \def\linewidth{0.4pt}  
    \tikzstyle{atom} = [draw, line width=\linewidth, circle, fill=white, anchor=north, inner sep=0.1pt, minimum size=0.6cm]
    \tikzstyle{conn} = [-{Latex}, line width=0.7pt]

    \begin{pgfonlayer}{foreground}
        \node[atom, line width=0.6pt] (a) at (0, 0) {$a$};
        \node[atom] (b) at (1.8, 0) {$b$};
        \node[atom] (c) at (0, 1.5) {$c$};
        \node[atom] (d) at (1.9, 1.5) {$d$};
    \end{pgfonlayer}

    \pic["$\varphi_{cab}$", draw, color=orange, angle eccentricity=0.7, angle radius=0.9cm, fill=orange!5] {angle = b--a--c};
    \pic["$\varphi_{abd}$", draw, color=blue, angle eccentricity=0.62, angle radius=0.9cm, fill=blue!5] {angle = d--b--a};

    \draw[conn] (c) -> (a) node[pos=0.35, above right, inner sep=0.5pt] (m_ca) {$\vm_{ca}$};
    \draw[] (b) -> (a);
    \draw[conn] (d) -> (b) node[pos=0.3, above left, inner sep=0.5pt] (m_db) {$\vm_{db}$};

    \draw[-{Latex},dashed] ([yshift=1.0pt]m_db.north west) to [out=135,in=45] ([shift=(0.5pt:1pt)]m_ca.north);


    \begin{pgfonlayer}{foreground}
        \node[atom, line width=0.6pt] (a2) at (3.1, 0) {$a$,$b$};
        \node[atom] (c2) at ($(a2.north) + (-0.4, 1.5)$) {$c$};
        \node[atom] (d2) at ($(a2.north) + (0.4, 1.5)$) {$d$};
    \end{pgfonlayer}

    \draw[{Latex[length=4pt]}-] ($0.5*(b) + 0.5*(a2) + (0, 0.1)$) arc [x radius=0.2cm, y radius=0.1cm, start angle=90, end angle=-250];
    \node[font=\scriptsize] (rot) at ($0.5*(b) + 0.5*(a2) - (0, 0.27)$) {(rotate)};

    \pic[draw, color=green, angle radius=0.9cm, fill=green!5] {angle = d2--a2--c2};

    \draw[conn] (c2) -> node[color=green, pos=0.6, left, inner sep=1pt] {$\theta_{cabd}$} (a2);
    \draw[conn] (d2) -> (a2);



\end{tikzpicture}
\endgroup
    \caption{Angles used in geometric message passing. The dihedral angle $\theta_{cabd}$ becomes visible when rotating the molecule so that atoms $a$ and $b$ lie on top of each other (right).}
    \label{fig:agg}
\end{wrapfigure}
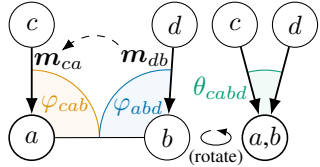

\textbf{Geometric representation.} We now develop a specific two-hop message passing scheme based on \cref{eq:gem}. We use embeddings based on interatomic directions, and embed all atom pairs with distance $x_{ca} \le c_{\text{emb}}$. $\hat{\vr}_o$ and $\hat{\vr}_i$ are thus instantiated as the interatomic directions $\hat{\vx}_{ca}$ and $\hat{\vx}_{db}$. We denote directional embeddings as $\vm_{ca} = \mH_a(\hat{\vx}_{ca})$. Message passing is thus based on quadruplets of atoms -- two atoms are interacting ($a$ and $b$) and two atoms define the directions ($c$ and $d$). We denote the angle between directions by $\varphi_{abd} = \measuredangle \hat{\vx}_{ab} \hat{\vx}_{db}$. To improve empirical performance we additionally use the dihedral angle $\theta_{cabd} = \measuredangle \hat{\vx}_{ca} \hat{\vx}_{db} \perp \hat{\vx}_{ba}$ and substitute $\measuredangle \hat{\vr}_o \hat{\vr}_i = \measuredangle \hat{\vx}_{ca} \hat{\vx}_{db}$ with $\varphi_{cab}$. \cref{fig:agg} illustrates the three angles $\varphi_{cab}$, $\varphi_{abd}$, and $\theta_{cabd}$ we use for updating the embedding $\vm_{ca}$ based on $\vm_{db}$. To ensure that all angles are well-defined we exclude overlapping atom quadruplets, i.e.\ $a\!\neq\!b\!\neq\!c\!\neq\!d$. We represent the relative directional information using spherical Fourier-Bessel bases with polynomial radial envelopes to ensure smoothly differentiable predictions, as proposed by \citet{gasteiger_directional_2020}. We split the basis into three parts to incorporate all available geometric information. Before the envelope, these are:
\begin{align}
    \tilde{\ve}_{\text{RBF}, n}(x_{db}) &= \sqrt{\frac{2}{c_{\text{emb}}}} \frac{\sin(\frac{n \pi}{c_{\text{emb}}} x_{db})}{x_{db}},\\
    \tilde{\ve}_{\text{CBF}, ln}(x_{ba}, \varphi_{abd}) &= \sqrt{\frac{2}{c_{\text{int}}^3 j_{l+1}^2(z_{ln})}} j_l(\frac{z_{ln}}{c_{\text{int}}} x_{ba}) Y_{l0}(\varphi_{abd}),\\
    \tilde{\ve}_{\text{SBF}, lmn}(x_{ca}, \varphi_{cab}, \theta_{cabd}) &= \sqrt{\frac{2}{c_{\text{emb}}^3 j_{l+1}^2(z_{ln})}} j_l(\frac{z_{ln}}{c_{\text{emb}}} x_{ca}) Y_{lm}(\varphi_{cab}, \theta_{cabd}),
\end{align}
with the interaction cutoff $c_{\text{int}}$, the spherical Bessel functions $j_l$, and the \mbox{$n$-th} root of the \mbox{$l$-order} Bessel function $z_{ln}$. Note that \citet{gasteiger_directional_2020} only used the first two parts $\ve_{\text{RBF}}$ and $\ve_{\text{CBF}}$. These representations are then transformed using two linear layers to obtain the filter $F$. In order to maintain a smoothly differentiable cutoff we cannot use a bias in this transformation. Altogether, the core geometric message passing scheme is
\begingroup
\setlength{\jot}{-3.5ex}
\begin{equation}
\begin{aligned}
    \tilde{\vm}_{ca} = \sum_{\substack{b \in \gN_a^{\text{int}} \setminus \{c\}, \\ d \in \gN_b^{\text{emb}} \setminus \{a,c\}}} \Big( &(\mW_{\text{SBF}1} \ve_{\text{SBF}}(x_{ca}, \varphi_{cab}, \theta_{cabd}))^T \tW ((\mW_{\text{CBF}2} \mW_{\text{CBF}1} \ve_{\text{CBF}}(x_{ba}, \varphi_{abd}))\\
    &\odot (\mW_{\text{RBF}2} \mW_{\text{RBF}1} \ve_{\text{RBF}}(x_{db})) \odot \vm_{db} )\Big),
\end{aligned}
\label{eq:core_geom}
\end{equation}
\endgroup
where $\mW$ denotes a weight matrix, $\tW$ denotes a weight tensor. The first weight matrix of each representation part has a small output dimension. This causes a bottleneck that improves generalization.

\textbf{Symmetric message passing.} Whenever we have a directional embedding $\vm_{ca}$, we also have the opposing embedding $\vm_{ac}$, since both are based on the same cutoff $c_{\text{emb}}$. Whether we associate the embedding $\vm_{ca}$ or $\vm_{ac}$ with atom $a$ is arbitrary. A more principled approach is to \emph{jointly} interpret both embeddings as a representation of the atom pair $a$ and $c$. In this view, an update to $\vm_{ca}$ should also influence $\vm_{ac}$. This would normally require executing the above message passing scheme twice, once for updating $\vm_{ca}$ based on $\vm_{db}$, and once for updating $\vm_{ac}$ based on $\vm_{db}$. We propose to circumvent this double execution by calculating the update (\cref{eq:core_geom}) only once and then using it for both $\vm_{ca}$ and $\vm_{ac}$. To preserve the distinction between the two directions and ensure that $\vm_{ca} \neq \vm_{ac}$, we transform the two updates using two separate learnable weight matrices. One single message passing update thus carries information for both embeddings, which is then dissected by the two weight matrices. In practice, this only requires a simple re-indexing operation that maps the edge $ca$ to $ac$.

\textbf{Efficient bilinear layer.} The whole message passing scheme, i.e.\ basis transformation, neighbor aggregation, and bilinear layer, only use linear functions. We can therefore freely optimize the order of summation without changing the result, as proposed by \citet{wu_pointconv_2019} (see \cref{app:eff_bilinear} for details). Doing so can provide a faster and more memory-efficient model, reducing memory usage by \SI{50}{\percent} even for Hadamard products. Moreover, since everything is based on efficient matrix products, this allows us to use the bilinear layer at practically no additional cost compared to a Hadamard product. Note that this requires using padded matrices instead of the usual gather-scatter operations to prevent excessively large intermediate results.

\section{GemNet: Geometric message passing neural network} \label{sec:gemnet}

\begin{figure}[t]
    \centering
    \resizebox{\textwidth}{!}{
    \input{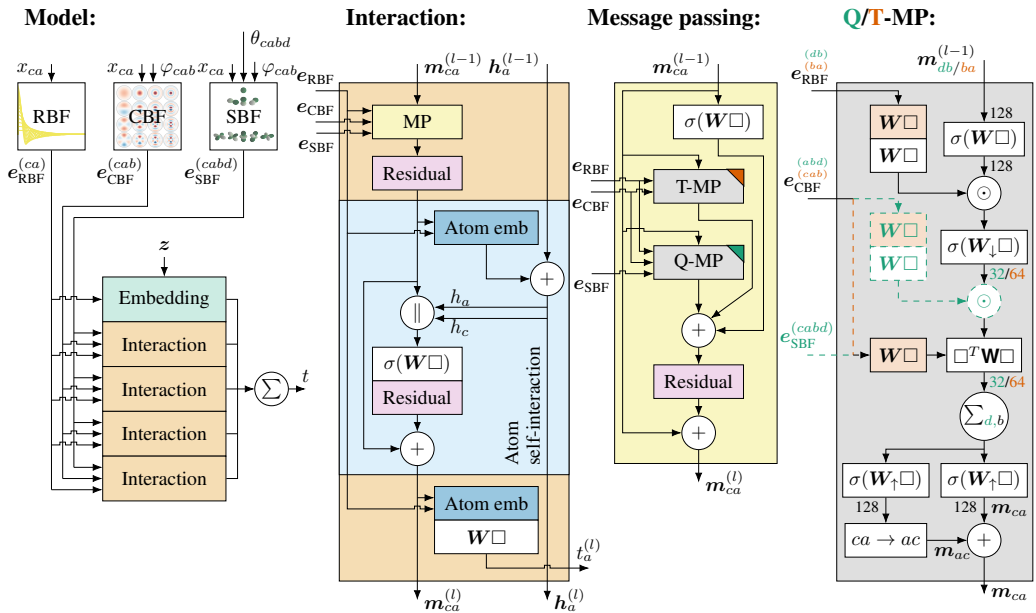}
    }
    \caption{The GemNet architecture (comprehensive version in \cref{app:gemnet}). $\square$ denotes the layer's input, $\|$ concatenation, and $\sigma$ a non-linearity. Directional embeddings $\vm_{ca}$ are updated using three forms of interaction: Two-hop geometric message passing (Q-MP), one-hop geometric message passing (T-MP), and atom self-interactions. Differences between Q-MP and T-MP are denoted by colors and dashed lines.}
    \label{fig:gemnet}
\end{figure}

\textbf{GemNet.} The geometric message passing neural network (GemNet) is a significantly refined architecture based on DimeNet$^{++}$ \citep[Hippocratic license 2.1]{gasteiger_fast_2020}. GemNet predicts the molecular energy $E$ and forces $\mF \in \R^{3 \times n}$ based on the atomic positions $\mX \in \R^{3 \times n}$ and the atomic numbers $\vz \in \sN^n$. The architecture is illustrated in \cref{fig:gemnet}. A comprehensive version with low-level layers and hyperparameters is described in \cref{app:gemnet}. GemNet was developed on the COLL dataset, but generalizes to other datasets such as MD17 without architectural changes. Every change we propose either improves model performance or reduces model complexity. For example, GemNet uses no biases since we found them to be irrelevant or even detrimental to accuracy. We show the impact of the most relevant changes via ablation studies in \cref{sec:exp}. \looseness=-1

\textbf{Interactions.} GemNet incorporates three forms of interactions. The first is geometric message passing, as described in \cref{sec:gmp}. The second is a one-hop form of geometric message passing. This interaction uses a single cutoff $c = c_{\text{emb}}$ and passes messages only between directional embeddings pointing towards the same atom, similarly to DimeNet \citep{gasteiger_directional_2020}. This provides both angle-based pair interactions and atom self-interactions, thanks to the symmetric message passing scheme described in \cref{sec:gmp}. The third interaction is a pure atom self-interaction based on atom embeddings. We first aggregate the directional embeddings of one atom to obtain an atom embedding. We then use this atom embedding to update all directional embeddings. We found all three interaction forms to be beneficial, and show this in our ablation studies.

\textbf{Stabilizing activation variance.} The variance of activations in a model is usually stabilized using normalization methods, which has various positive effects on training \citep{de_batch_2020,santurkar_how_2018,luo_towards_2019}. However, they also have multiple undesirable side effects, especially in the context of molecular regression. Batch normalization introduces correlations between separate molecules and atoms. Layer normalization forces all activation scales to be constant, while atomic interactions actually cover a large range of scales -- directly bonding atoms have a substantially stronger interaction than atoms at a long range. To circumvent these issues, we stabilize GemNet's variance by introducing constant scaling factors, as suggested by \citet{brock_characterizing_2021}. We found that the activation variance is primarily impacted by four components: Skip connections, non-linearities, message aggregation, and Hadamard/bilinear layers. The two summands in a skip connection $y = x + f(x)$ have no covariance at initialization due to random weight matrices. We can thus remove its impact by scaling the output by $1/\sqrt{2}$. We remove the non-linearity's impact by scaling its output with a gain of $\gamma = 1/0.6$ for SiLU, similarly to \citep{klambauer_self-normalizing_2017}. Note that we do not center SiLU's output but instead choose a slightly lower $\gamma$ to account for mean shift. Additionally, we standardize the weight matrices to have exactly zero mean and $1/\text{fan-in}$ variance. The sum aggregation and Hadamard/bilinear layers have a more complex impact on the variance, which we cannot determine a priori (see \cref{app:var_mp} for details). We therefore estimate the variance after these layers based on random batches of data. We then rescale their output accordingly to obtain roughly the variance of the layer input at initialization. These simple empirical scaling factors are sufficient to keep the activation variance roughly constant (see \cref{fig:var_scale}). We found that other measures such as adaptive gradient clipping \citep{brock_high-performance_2021}, scaled weight standardization \citep{brock_characterizing_2021}, or weighting the residual block with zero at initialization \citep{de_batch_2020} are not beneficial for model accuracy.

\textbf{GemNet-Q and GemNet-T.} Geometric message passing is comparatively expensive since it is based on quadruplets of atoms. Its runtime thus scales with $\gO(nk_{\text{int}}k_{\text{emb}}^2)$, where $k_{\text{int}}$ is the number of interacting neighbors, and $k_{\text{emb}}$ is the number of embedded directions. For this reason we investigate two message passing models in our experiments -- one with two-hop geometric message passing (GemNet-Q) and one using only the two cheaper forms of interaction (GemNet-T). Their complexities are $\gO(nk_{\text{int}}k_{\text{emb}}^2)$ and $\gO(nk_{\text{emb}}^2)$, respectively. Note that GemNet-T is thus a direct ablation of the two-hop message passing scheme implied by our theoretical results.

\textbf{Direct force predictions.} GemNet predicts forces by calculating $\mF_a = - \partial E / \partial \vx_a$ via backpropagation. This form of calculation guarantees a conservative force field, which is important for the stability of simulations. However, by using \cref{eq:vec_pred} we can also directly predict forces and other vector quantities. This essentially means predicting a magnitude for each directional embedding and then summing up over the vectors defined by this magnitude and the embedding's associated direction, similarly to \citet{park_accurate_2021}. We denote this variant as \emph{GemNet-dQ} and \emph{GemNet-dT}. Interestingly, GemNet is thus able to generate rotationally \emph{equivariant} predictions despite only using \emph{invariant} representations. Direct predictions substantially accelerate the model, especially for training. For most datasets, the resulting accuracy is on par with most previous models, but significantly worse than GemNet's accuracy via backpropagation. However, this is not true for OC20, where we found GemNet-dT to converge faster and perform on par with GemNet-T.

\textbf{Limitations.} GemNet is focused on one specific, important task: Predictions for molecular simulations. We do not make any statements regarding its performance beyond this scope. The GemNet architecture might seem more complex than some previous models, due to its larger variety of interactions and blocks. However, its number of parameters and training or inference time is actually on par with previous models. Two-hop message passing introduces significant computational overhead. We mitigate this effect with a down-projection layer and additionally introduce the ablated GemNet-T model. This model performs surprisingly well on MD17, but not on COLL. This suggests that one-hop message passing is expressive enough for some practical use cases, but two-hop message passing gives an advantage for the more challenging task of fitting multiple molecules at once.

\textbf{Societal impacts.} Accelerating molecular simulations can have positive effects in a wide range of applications in physics and chemistry. At the same time, however, this can be used for malicious purposes such as developing chemical agents or weapons. To the best of our knowledge, this work does not promote these use cases more than regular chemistry research does. To somewhat mitigate negative effects we will publish our source code under the Hippocratic license \citep{ehmke_hippocratic_2020}.

\section{Experiments} \label{sec:exp}

\begin{table}
    \centering
    \begin{minipage}[t]{0.38\textwidth}
        \centering
        \caption{MAE on \textsc{COLL}, in \si[per-mode=symbol]{\milli\electronvolt\per\angstrom} and \si{\milli\electronvolt}. GemNet is \SI{34}{\percent} more accurate for forces. The higher energy error is due to its lower loss weight.}
        \begin{tabular}{lS[table-format=3.1]@{\hspace{0.2cm}}S[table-format=3]}
& \mcc{Forces}              & \mcc{Energy}            \\
\hline
SchNet                & 172           & 198         \\
$\text{DimeNet}^{++}$ & 40            & \bfseries 47 \\
GemNet-Q              & \bfseries 26.4 & 53         \\
GemNet-T              & 31.6          & 60          \\
GemNet-dQ             & 38.1          & 60          \\
GemNet-dT             & 43.1          & 55
\end{tabular}

        \label{tab:coll}
    \end{minipage}
    \hfill
    \begin{minipage}[t]{0.6\textwidth}
        \centering
        \caption{Force MAE for MD17@CCSD in \si[per-mode=symbol]{\milli\electronvolt\per\angstrom}. GemNet outperforms previous methods by \SI{44}{\percent} on average.}
        \begin{tabular}{lS@{\hspace{0.2cm}}S@{\hspace{0.2cm}}S@{\hspace{0.2cm}}SS}
               &      \mcc{sGDML}     &     \mcc{NequIP}    &   \mcc{GemNet-Q}  & \mcc{GemNet-T} \\
\hline
Aspirin        &   33.0  & 14.7   &   10.4   &   \bfseries 10.3   \\
Benzene        &    1.7  &  0.8   &   \bfseries 0.7   &   \bfseries 0.7   \\
Ethanol        &   15.2  &  9.4   &   \bfseries 3.1   &   \bfseries 3.1   \\
Malonaldehyde  &   16.0  & 16.0   &    6.0   &   \bfseries 5.9   \\
Toluene        &    9.1  &  4.4   &   \bfseries 2.5   &    2.7   \\
\end{tabular}

        \label{tab:ccsd}
    \end{minipage}
\end{table}

\begin{table}
    \centering
    \caption{Force MAE for MD17 in \si[per-mode=symbol]{\milli\electronvolt\per\angstrom}. GemNet outperforms all previous methods by a wide margin, on average by \SI{41}{\percent}.}
    \begin{tabular}{lSSSSSSSSS[table-format=1.1]S[table-format=1.1]}
               &  \multicolumn{2}{c}{Kernel methods}       & &   \multicolumn{4}{c}{GNNs}    & & \multicolumn{2}{c}{\textbf{GemNet}} \\
\cline{2-3} \cline{5-8} \cline{10-11}
                                    & \mcc{sGDML \rule{0pt}{0.9em}}        &   \mcc{FCHL19}       & &   \mcc{DimeNet}      &   \mcc{SphereNet}    &   \mcc{NequIP}       &   \mcc{PaiNN}       & &   \mcc{GemNet-Q}     &   \mcc{GemNet-T}              \\
\hline
Aspirin \rule{0pt}{0.9em}                             & 29.5   &   20.7   & &   21.6   &   18.6   &   15.1   &   14.7  & &   \bfseries 9.4   &    9.5   \\
Benzene\citep{chmiela_machine_2017} &      \mcc{-}  &        \mcc{-}  & &    8.1   &    7.7   &    8.1   &        \mcc{-} & &   \bfseries 6.3   &   \bfseries 6.3   \\
Benzene\citep{chmiela_towards_2018} &  2.6   &        \mcc{-}  & &        \mcc{-}  &        \mcc{-}  &    2.3   &        \mcc{-} & &    1.5   &   \bfseries 1.4   \\
Ethanol                             & 14.3   &    5.9   & &   10.0   &    9.0   &    9.0   &    9.7  & &    3.8   &   \bfseries 3.7   \\
Malonaldehyde                       & 17.8   &   10.6   & &   16.6   &   14.7   &   14.6   &   14.9  & &    6.9   &   \bfseries 6.7   \\
Naphthalene                         &  4.8   &    6.5   & &    9.3   &    7.7   &    4.2   &    3.3  & &   \bfseries 2.2   &    2.4   \\
Salicylic acid                      & 12.1   &    9.6   & &   16.2   &   15.6   &   10.3   &    8.5  & &   \bfseries 5.4   &    5.5   \\
Toluene                             &  6.1   &    8.8   & &    9.4   &    6.7   &    4.4   &    4.1  & &   \bfseries 2.6   &   \bfseries 2.6   \\
Uracil                              & 10.4   &    4.6   & &   13.1   &   11.6   &    7.5   &    6.0  & &    4.5   &   \bfseries 4.2
\end{tabular}

    \label{tab:md17}
\end{table}

\begin{table}
    \centering
    \caption{Results for the three tasks of the open catalyst dataset (OC20), averaged across its four test sets. GemNet outperforms all previous methods in all measures, on average by \SI{20}{\percent}.\\
    {\footnotesize *DimeNet$^{++}$-large uses separate models for energy and force prediction for IS2RE.}}
    \begin{tabular}{lS[table-format=3.1]SS[table-format=1.3]cSScS[table-format=3.1]}
                                 & \multicolumn{3}{c}{S2EF}                                                                                                                  &  & \multicolumn{2}{c}{IS2RS}                                       &  & \mcc{IS2RE} \\ \cline{2-4} \cline{6-7} \cline{9-9}
                                 & \mcc{Energy MAE \rule{0pt}{0.9em}}          & \mcc{Force MAE}                                                           & \mcc{Force cos}  &  & \mcc{AFbT}                     & \mcc{ADwT}                     &  & \mcc{Energy MAE}                            \\
                                 & \mcc{\si{\milli\electronvolt} $\downarrow$} & \mcc{\si[per-mode=symbol]{\milli\electronvolt\per\angstrom} $\downarrow$} & \mcc{$\uparrow$} &  & \mcc{\si{\percent} $\uparrow$} & \mcc{\si{\percent} $\uparrow$} &  & \mcc{\si{\milli\electronvolt} $\downarrow$} \\ \hline
ForceNet-large \rule{0pt}{0.9em} & \mcc{-}                                     & 31.2                                                                      & 0.520            &  & 12.7                           & 49.6                           &  & \mcc{-}                                     \\
DimeNet$^{++}$-large*            & \mcc{-}                                     & 31.3                                                                      & 0.544            &  & 21.8                           & 51.7                           &  & 559.1                                       \\
SpinConv                         & 336.3                                       & 29.7                                                                      & 0.539            &  & 16.7                           & 53.6                           &  & 434.3                                       \\
\textbf{GemNet-dT}               & \bfseries 292.4                             & \bfseries 24.2                                                            & \bfseries 0.616  &  & \bfseries 27.6                 & \bfseries 58.7                 &  & \bfseries 399.7
\end{tabular}

    \label{tab:oc20}
\end{table}


\textbf{Experimental setup.} We evaluate our model on four molecular dynamics datasets. \textsc{COLL} \citep[CC-BY 4.0]{gasteiger_fast_2020} consists of configurations taken from molecular collisions of different small organic molecules. MD17 \citep{chmiela_machine_2017} consists of configurations of multiple separate, thermalized molecules, considering only one molecule at a time. MD17@CCSD \citep{chmiela_towards_2018} uses the same setup, but calculates the forces using the more accurate and expensive CCSD or CCSD(T) method. The open catalyst (OC20)  dataset \citep[CC-BY 4.0]{chanussot_open_2021} consists of energy relaxation trajectories of solid catalysts with adsorbate molecules. This dataset is split into three tasks: (1) Structure to energy and forces (S2EF), which is the same task as used by the \textsc{COLL} and MD17 datasets, (2) initial structure to relaxed structure (IS2RS), where an energy optimization is carried out based on the model's predictions and we measure how close the final structure is to the true relaxed structure (average distance within threshold, ADwT) and whether the final forces are close to zero (average forces below threshold, AFbT), and (3) initial structure to relaxed energy (IS2RE), where we predict the energy of the relaxed structure, based on an energy optimization starting at the initial structure. All presented OC20 models are trained on the S2EF data.
Following the setup of \citet{batzner_se3-equivariant_2021}, we use 1000 training and validation configurations for MD17, and 950 training and 50 validation configurations for MD17@CCSD. We focus on force predictions and use a high force loss weight since they determine the accuracy of molecular simulations. We measure the mean absolute error (MAE), averaged over all samples, atoms, and components. We compare with the results reported by several state-of-the-art models: sGDML \citep{chmiela_towards_2018}, FCHL19 \citep{christensen_fchl_2020}, SchNet \citep{schutt_schnet:_2017}, DimeNet \citep{gasteiger_directional_2020}, DimeNet$^{++}$ \citep{gasteiger_fast_2020}, SphereNet \citep{liu_spherical_2021}, NequIP \citep{batzner_se3-equivariant_2021}, PaiNN \citep{schutt_equivariant_2021}, ForceNet \citep{hu_forcenet_2021}, and SpinConv \citep{shuaibi_rotation_2021}. For further details see \cref{app:training}.

\textbf{Results.} \cref{tab:coll,tab:ccsd,tab:md17,tab:oc20} show that GemNet-T and GemNet-Q consistently perform best on all molecular dynamics datasets investigated -- and by a large margin. This is true both in comparison to previous GNNs and for kernel methods -- despite the latter typically being more sample efficient. The improvements are largest for chain-like molecules, such as ethanol and malonaldehyde. These molecules are the most challenging since they exhibit a wide range of movement.
GemNet even performs better than some previous models that were trained with 50x more training samples. For example, it performs better than SchNet with \num{50000} training samples on six out of eight MD17 molecules (see \cref{tab:md17_50k}).
Interestingly, the two-hop message passing scheme implied by our theoretical results (GemNet-Q) yields significant improvements on \textsc{COLL}, but performs approximately on par with the ablated GemNet-T on MD17. To investigate this disagreement we trained GemNet on a combined dataset of all MD17 molecules. \cref{tab:md17_combined} shows that GemNet-Q again performs better than GemNet-T in this setting. These results suggest that regular MD17 is too simple to show the benefits of two-hop message passing. It seems to be particularly important in more difficult settings that cover a large variety of configurations and molecules.

\textbf{Computational aspects.} GemNet-Q is roughly two times slower than GemNet-T (see \cref{tab:runtimes}). Thanks to the efficient aggregation, GemNet with bilinear layers is as fast as with regular Hadamard products. Efficient aggregation also reduces the memory usage for regular Hadamard products by around \SI{50}{\percent} (from 4.1GB to 2.2GB for a batch of 32 Toluene molecules). Note that GemNet has not been optimized for runtime and can likely be accelerated substantially. GemNet-Q uses 2.2M and GemNet-T 1.9M parameters, which is comparable to previous models such as DimeNet$^{++}$, which uses 1.9M parameters. See \cref{app:comp_time} for further details.

\textbf{Direct force prediction.} Directly predicting the forces accelerates training by four times on average and inference by 1.6 times on average in our experiments (see \cref{tab:runtimes}), while reducing memory consumption by roughly a factor of two. While using direct predictions instead of backpropagation increases the MAE by \SI{44}{\percent} on \textsc{COLL} and by \SI{48}{\percent} on MD17 (see \cref{tab:coll,tab:direct}), they actually perform better on the S2EF task on OC20. This is likely due to OC20 being orders of magnitude larger than \textsc{COLL} and MD17. Whether to use direct predictions thus depends on the dataset and the application's computational requirements.

\begin{wraptable}[16]{r}{0.52\textwidth}
    \centering
    \vspace*{-0.3cm}
    \caption{Ablation studies on \textsc{COLL}. Force MAE in \si[per-mode=symbol]{\milli\electronvolt\per\angstrom} after \num{500000} training steps. All proposed components yield significant improvements.}
    \begin{tabular}{lS}
Model         &    \mcc{Forces} \\
\hline
without symmetric message passing                 &  28.5  \\
Hadamard product instead of bilinear layer        &  29.3  \\
\hline
without atom embedding updates                    &  28.3  \\
without one-hop message passing                   &  31.3  \\
without two-hop message passing                   &  32.4  \\
\hline
without scaling factors                           &  29.1  \\
use layer norm instead (without centering)        &  33.3  \\
\hline
with bias                                         &  27.2  \\
GemNet-Q                                          &  27.0  \\
\end{tabular}

    \label{tab:ablation}
\end{wraptable}

\textbf{Ablation studies.} We investigate the proposed architectural improvements on \textsc{COLL} in \cref{tab:ablation}. The proposed symmetric message passing scheme yields significant accuracy improvements, as does using a bilinear layers instead of a Hadamard product. We also see that removing any of the three interaction forms described in \cref{sec:gemnet} increases the error, showing that this combination is indeed beneficial. The proposed scaling factors also yield decent improvements, while regular layer normalization actually increases the error. Two-hop message passing yields the largest single improvement. \cref{tab:ablation_dimenet} shows that our architectural improvements yield similar benefits for DimeNet$^{++}$. Overall, the error improvements are quite evenly distributed. This suggests that GemNet's improved performance is not due to one single change, but rather due to the full range of improvements proposed in this work.

\section{Conclusion}

In this work we proved the universality for GNNs using spherical representations. We proposed geometric message passing based on these insights, and improved this scheme with symmetric message passing and efficient bilinear layers. We incorporated these improvements in the GemNet architecture, which substantially improves the error on various molecular dynamics datasets. We showed that all of the proposed enhancements yield significant performance improvements. Most of our proposed improvements are of independent interest for other molecular GNNs.

\begin{ack}
We would like to thank Soledad Villar for help with proving \cref{lem:perminv_vec}, as well as Nicholas Gao and Aleksandar Bojchevski for their invaluable feedback and comments.

This research was supported by the Deutsche Forschungsgemeinschaft (DFG) through the Emmy Noether grant GU 1409/2-1 and the TUM International Graduate School of Science and Engineering (IGSSE), GSC 81.
\end{ack}

{
\small
\bibliography{library}
\bibliographystyle{gasteiger}
}

\appendix
\setcounter{theorem}{0}
\renewcommand*{\thetheorem}{\Alph{theorem}}
\setcounter{lemma}{0}
\renewcommand*{\thelemma}{\Alph{lemma}}
\setcounter{proposition}{0}
\renewcommand*{\theproposition}{\Alph{proposition}}

\section{Proof of \cref{th:univ_sphere}} \label{app:univ_sphere}
We prove the universal approximation theorem by showing the equivalence of TFN and our model. Complex spherical harmonics are related to Clebsch-Gordan coefficients via \citep[3.7.72]{sakurai_modern_1993}
\begin{equation}
    Y_{m_i}^{(l_i)}(\hat{\vr}) Y_{m_f}^{(l_f)}(\hat{\vr}) = \sum_{l_o, m_o} \sqrt{\frac{(2l_i + 1)(2l_f + 1)}{4\pi (2 l_o + 1)}} C_{(l_f, 0), (l_i, 0)}^{(l_o, 0)} C_{(l_f, m_f), (l_i, m_i)}^{(l_o, m_o)} Y_{m_o}^{(l_o)}(\hat{\vr}).
\end{equation}
We now use the fact that multiplying a learnable function with a unitary matrix or a scalar does not change the resulting function space. We can therefore adapt \cref{eq:tfn} by substituting
\begin{equation}
    C_{(l_f, m_f), (l_i, m_i)}^{(l_o, m_o)} \mapsto C(l_f, m_f, l_i, m_i, l_o, m_o) = \sqrt{\frac{(2l_i + 1)(2l_f + 1)}{4\pi (2 l_o + 1)}} C_{(l_f, 0), (l_i, 0)}^{(l_o, 0)} C_{(l_f, m_f), (l_i, m_i)}^{(l_o, m_o)}
\end{equation}
without impacting model expressivity. Since real spherical harmonics and complex (conjugate) spherical harmonics cover the same function space, we can furthermore substitute the filter with $F'^{(l)}_{m}(\vx) = R^{(l)}(x) Y^{(l)*}_{m}(\hat{\vx})$. Using the spherical harmonics expansion we therefore obtain
\begin{equation}
\begin{aligned}
    \tilde{\mH}'_a &(\mX, \mH')(\hat{\vr}) = \sum_{l_o, m_o} \tilde{\mH}'^{(l_o)}_{am_o}(\mX, \mH') Y_{m_o}^{(l_o)}(\hat{\vr})\\
    = {}& \sum_{l_o, m_o} \left( \theta \mH'^{(l_o)}_{am_o} + \sum_{l_f, m_f} \sum_{l_i, m_i} C(l_f, m_f, l_i, m_i, l_o, m_o) \sum_{b \in \gN_a} F'^{(l_f)}_{m_f}(\vx_{ba}) \mH'^{(l_i)}_{bm_i} \right) Y_{m_o}^{(l_o)}(\hat{\vr})\\
    = {}& \theta \mH'_a(\hat{\vr}) + \sum_{l_f, m_f} \sum_{l_i, m_i} \sum_{b \in \gN_a} F'^{(l_f)}_{m_f}(\vx_{ba}) Y_{m_f}^{(l_f)}(\hat{\vr}) \mH'^{(l_i)}_{bm_i} Y_{m_i}^{(l_i)}(\hat{\vr})\\
    = {}& \theta \mH'_a(\hat{\vr}) + \sum_{b \in \gN_a} \left( \sum_{l_f, m_f} F'^{(l_f)}_{m_f}(\vx_{ba}) Y_{m_f}^{(l_f)}(\hat{\vr}) \right) \left( \sum_{l_i, m_i} \mH'^{(l_i)}_{bm_i} Y_{m_i}^{(l_i)}(\hat{\vr}) \right)\\
    = {}& \theta \mH'_a(\hat{\vr}) + \sum_{b \in \gN_a} F'(\vx_{ba}, \hat{\vr}) \mH'_b(\hat{\vr}).
\end{aligned}
\end{equation}
These functions rely on complex-valued representations, while the output and $\SO3$ representations are real-valued. However, we can restrict the representations to real values without changing the resulting function space. To see this, we look at the result's real component
\begin{equation}
\begin{aligned}
    \Re[\tilde{\mH}'_a (\mX, \mH')(\hat{\vr})] &= \theta \Re[\mH'_a(\hat{\vr})] + \sum_{b \in \gN_a} \Re[F'(\vx_{ba}, \hat{\vr}) \mH'_b(\hat{\vr})]\\
    &= \theta \Re[\mH'_a(\hat{\vr})] + \sum_{b \in \gN_a} (\Re[F'(\vx_{ba}, \hat{\vr})] \Re[\mH'_b(\hat{\vr})] - \Im[F'(\vx_{ba}, \hat{\vr})] \Im[\mH'_b(\hat{\vr})]).
\end{aligned}
\end{equation}
The function space covered by $\Re[F'(\vx, \hat{\vr})]$, and thus $\Re[\mH'(\hat{\vr})]$, is the same as $\Im[F'(\vx, \hat{\vr})]$, and thus $\Im[\mH'(\hat{\vr})]$. We can therefore simply remove the imaginary part without changing the resulting function space, obtaining
\begin{equation}
\begin{aligned}
    \tilde{\mH}^{\text{sphere}}_a (\mX, \mH)(\hat{\vr}) &= \theta \mH_a(\hat{\vr}) + \sum_{b \in \gN_a} \Re[F'(\vx_{ba}, \hat{\vr})] \mH_b(\hat{\vr})\\
    &= \theta \mH_a(\hat{\vr}) + \sum_{b \in \gN_a} F_{\text{sphere}}(\vx_{ba}, \hat{\vr}) \mH_b(\hat{\vr}).
\end{aligned}
\end{equation}
$\gF_{\text{feat}}^{\text{sphere}}(D)$ thus spans the exact same space of embedding functions as $\gF_{\text{feat}}^{\text{TFN}}(D)$, despite only using real functions on the $S^2$ sphere. However, we cannot span the full space of rotationally equivariant linear pooling functions, since equivariant linear functions on the $S^2$ sphere are limited to convolutions with zonal filters \citep{esteves_learning_2018}. Fortunately, scalar pooling functions are limited to linear functions of the constant $l=0$ part. This is equivalent to integrating over the real-space spherical representation, as done in $\gF_{\text{pool}}^{\text{sphere}}$.\qed

\section{Proof of \cref{th:univ_vec}} \label{app:univ_vec}

To prove this theorem we first introduce a proposition by \citet{villar_scalars_2021}.

\begin{proposition}[\citet{villar_scalars_2021}]
    If $h$ is an $\SOg(d)$-equivariant function $\R^{d \times n} \to \R^d$ of $n$ vector inputs $\vx_1, \vx_2, \ldots, \vx_n$, then there are $n$ $\SOg(d)$-invariant functions $f_c \colon \R^{d \times n} \to \R$ such that
    \begin{equation}
        h(\vx_1, \vx_2, \ldots, \vx_n) = \sum_{c=1}^n f^{(c)}(\vx_1, \vx_2, \ldots, \vx_n) \vx_c,
    \end{equation}
    except when $\vx_1, \vx_2, \ldots, \vx_n$ span a $(d - 1)$-dimensional space. In that case, there exist $\Og(d)$-invariant functions $f_c \colon \R^{d \times n} \to \R$ such that
    \begin{equation}
        h(\vx_1, \vx_2, \ldots, \vx_n) = \sum_{c=1}^n f^{(c)}(\vx_1, \vx_2, \ldots, \vx_n) \vx_c + \sum_{S \in {[n] \choose d-1}} f^{(S)}(\vx_1, \vx_2, \ldots, \vx_n) \vx_S,
    \end{equation}
    where $[n] := \{1,\ldots,n\}$, ${[n] \choose d-1}$ is the set of all $(d - 1)$-subsets of $[n]$, and $\vx_S$ is the generalized cross product of vectors $\vx_i$ with $i \in S$ (taken in ascending order).
    \label{prop:so_vec}
\end{proposition}

To extend \cref{prop:so_vec} to our case, we need to restrict the functions to being translation-invariant and permutation-equivariant. We will only concern ourselves with the case where the vectors do not span a $(d - 1)$-dimensional space. We start by considering translation-invariant functions, following the proof idea of \citet[Lemma 7]{villar_scalars_2021}.

\begin{lemma}
    Let $h$ be a translation-invariant and $\SOg(d)$-equivariant function $\R^{d \times n} \to \R^d$ of $n$ vector inputs $\vx_1, \vx_2, \ldots, \vx_n$. Let $\vx_2 - \vx_1, \ldots, \vx_n - \vx_1$ not span a $(d - 1)$-dimensional space. Then there are $n - 1$ translation- and $\SOg(d)$-invariant functions $f_c \colon \R^{d \times n} \to \R$ such that
    \begin{equation}
        h(\vx_1, \vx_2, \ldots, \vx_n) = \sum_{c=2}^n f^{(c)}(\vx_1, \vx_2, \ldots, \vx_n) (\vx_c - \vx_1).
    \end{equation}
    \label{lem:transinv_vec}
\end{lemma}
\emph{Proof.} Consider the $\SOg(d)$-equivariant function $\tilde{h} \colon \R^{d \times (n-1)} \to \R^d$ with
\begin{equation}
    h(\vx_1, \vx_2, \ldots, \vx_n) = h(0, \vx_2 - \vx_1, \ldots, \vx_n - \vx_1) = \tilde{h}(\vx_2 - \vx_1, \ldots, \vx_n - \vx_1).
\label{eq:h_transinv}
\end{equation}
Due to \cref{prop:so_vec} we have
\begin{equation}
    \tilde{h}(\vx_2 - \vx_1, \ldots, \vx_n - \vx_1) = \sum_{c=2}^n \tilde{f}^{(c)}(\vx_2 - \vx_1, \ldots, \vx_n - \vx_1) (\vx_c - \vx_1),
\end{equation}
with the $\SOg(d)$-equivariant function $\tilde{f}^{(c)}$. If we now substitute $\tilde{f}^{(c)}$ with the $\SOg(d)$-equivariant and translation-invariant function $f^{(c)}$, i.e.
\begin{equation}
    \tilde{f}^{(c)}(\vx_2 - \vx_1, \ldots, \vx_n - \vx_1) = f^{(c)}(0, \vx_2 - \vx_1, \ldots, \vx_n - \vx_1) = f^{(c)}(\vx_1, \vx_2, \ldots, \vx_n),
\end{equation}
we obtain
\begin{equation}
    h(\vx_1, \vx_2, \ldots, \vx_n) = \sum_{c=2}^n f^{(c)}(\vx_1, \vx_2, \ldots, \vx_n) (\vx_c - \vx_1).
\end{equation} \qed

Next, we extend this result to permutation-equivariant functions.

\begin{lemma}
    Let $h$ be a translation-invariant, and permutation and $\SOg(d)$-equivariant function $\R^{d \times n} \to \R^{d \times n}$ of $n$ vector inputs $\vx_1, \vx_2, \ldots, \vx_n$. Let $\vx_2 - \vx_1, \ldots, \vx_n - \vx_1$ not span a $(d - 1)$-dimensional space. Then there are $n - 1$ translation- and $\SOg(d)$-invariant, and permutation-equivariant functions $f_c \colon \R^{d \times n} \to \R^{n}$ such that
    \begin{equation}
        h(\vx_1, \vx_2, \ldots, \vx_n) = \sum_{c=2}^n f^{(c)}(\vx_1, \vx_2, \ldots, \vx_n) (\vx_c - \vx_1).
    \end{equation}
    \label{lem:perminv_vec}
\end{lemma}
\emph{Proof.} Permutation equivariance implies that for all $s$ and $t$ (w.l.o.g.\ $s < t$)
\begin{equation}
    h_s(\ldots, \vx_s, \ldots, \vx_t, \ldots) = h_t(\ldots, \vx_t, \ldots, \vx_s, \ldots).
\end{equation}
Due to \cref{lem:transinv_vec} we have
\begin{align}
    h_s(\ldots, \vx_s, \ldots, \vx_t, \ldots) &= \sum_{c=2}^n f_s^{(c)}(\ldots, \vx_s, \ldots, \vx_t, \ldots) (\vx_c - \vx_1),\\
    = h_t(\ldots, \vx_t, \ldots, \vx_s, \ldots) &= \sum_{c=2}^n f_t^{(c)}(\ldots, \vx_t, \ldots, \vx_s, \ldots) (\vx_c - \vx_1),
\end{align}
with $n - 1$ $\SOg(d)$- and translation-invariant functions $f^{(c)} \colon \R^{d \times n} \to \R^{n}$. We can solve this equation by choosing
\begin{equation}
    f_s^{(c)}(\ldots, \vx_s, \ldots, \vx_t, \ldots) = f_t^{(c)}(\ldots, \vx_t, \ldots, \vx_s, \ldots),
\end{equation}
i.e.\ permutation-equivariant functions $f^{(c)}$. \qed

Finally, to bring \cref{lem:perminv_vec} to the form presented in the theorem, we first observe that adding scalar inputs $\mH$ does not affect the proofs in this section. Second, we observe that subtracting by $\vx_1$ in \cref{eq:h_transinv} is arbitrary. To bring this more in line with GNNs we can instead subtract the input of each $h_a$ by $\vx_a$. This yields
\begin{equation}
    h_a(\mX, \mH) = \sum_{\substack{c=1 \\ c \neq a}}^n f_a^{(c)}(\mX, \mH) (\vx_c - \vx_a).
\end{equation} \qed

\section{Proof of \cref{lem:filter_inv}} \label{app:filter_inv}

Using the fact that the Wigner-D matrix is unitary, we obtain for any rotation matrix $\mR$:
\begin{equation}
\begin{aligned}
    F_{\text{sphere}}(\mR \vx, \mR \hat{\vr}) &= \sum_{l, m} R^{(l)}(x) \Re[Y_m^{(l)*}(\mR \hat{\vx}) Y_{m}^{(l)}(\mR \hat{\vr})]\\
    &= \sum_{l, m, m', m''} R^{(l)}(x) \Re[Y_{m'}^{(l)*}(\hat{\vx}) D_{m, m'}^{(l)*}(\mR) D_{m, m''}^{(l)}(\mR) Y_{m''}^{(l)}(\hat{\vr})]\\
    &= \sum_{l, m', m''} R^{(l)}(x) \Re[Y_{m'}^{(l)*}(\hat{\vx}) \delta_{m', m''} Y_{m''}^{(l)}(\hat{\vr})]\\
    &= \sum_{l, m'} R^{(l)}(x) \Re[Y_{m'}^{(l)*}(\hat{\vx}) Y_{m'}^{(l)}(\hat{\vr})] = F_{\text{sphere}}(\vx, \hat{\vr}).
\end{aligned}
\end{equation} \qed

\section{Efficient message passing} \label{app:eff_bilinear}

For clarity we demonstrate how to optimize the summation order using the simpler one-hop message passing. For a regular Hadamard product we reorder the sums as
\begin{equation}
\begin{aligned}
    \vm_{(ca)i} &= \sum_{b \in \gN_a \setminus \{c\}} \Big( \mW^{(2)} \mW^{(1)} \ve_{\text{CBF}}(x_{ca}, \varphi_{bac}) \Big)_i \vm_{(ba)i} \\
    &= \sum_{b \in \gN_a \setminus \{c\}} \Big( \sum_{j} \sum_{l} \sum_{n} \mW_{ij}^{(2)} \mW_{j(ln)}^{(1)} \ve_{\text{CBF}}^{\text{rad}}(x_{ca})_{ln} \ve_{\text{CBF}}^{\text{SH}}(\varphi_{bac})_{l}\Big) \vm_{(ba)i} \\
    &= \sum_{j} \mW_{ij}^{(2)} \sum_{l} \Big( \sum_{n} \mW_{j(ln)}^{(1)} \ve_{\text{CBF}}^{\text{rad}}(x_{ca})_{ln}\Big) \Big( \sum_{b \in \gN_a \setminus \{c\}} \ve_{\text{CBF}}^{\text{SH}}(\varphi_{bac})_{l}  \vm_{(ba)i}  \Big).
    \label{eq:hadamard}
\end{aligned}
\end{equation}
For a bilinear layer we use
\begin{equation}
\begin{aligned}
    \vm_{(ca)i} &= \sum_{b \in \gN_a \setminus \{c\}} \Big( \Big( \mW^{(1)} \ve_{\text{CBF}}(x_{ca}, \varphi_{bac}) \Big)^T \tW^{(2)} \vm_{(ba)}\Big)_i \\
    &= \sum_{b \in \gN_a \setminus \{c\}}\sum_{i'} \sum_{j} \Big( \sum_{l} \sum_{n} \mW_{j(ln)}^{(1)} \ve_{\text{CBF}}^{\text{rad}}(x_{ca})_{ln} \ve_{\text{CBF}}^{\text{SH}}(\varphi_{bac})_{l}\Big) \tW_{iji'}^{(2)} \vm_{(ba)i'} \\
    &= \sum_{j} \sum_{i'} \tW_{iji'}^{(2)} \sum_{l} \Big( \sum_{n} \mW_{j(ln)}^{(1)} \ve_{\text{CBF}}^{\text{rad}}(x_{ca})_{ln} \Big) \Big( \sum_{b \in \gN_a \setminus \{c\}} \ve_{\text{CBF}}^{\text{SH}}(\varphi_{bac})_{l}  \vm_{(ba)i'}  \Big).
    \label{eq:bilinear}
\end{aligned}
\end{equation}
Note that since $\mW^{(1)}$ is shared across layers we only need to calculate the sum over $n$ once.

\section{Variance after message passing} \label{app:var_mp}

The layer-wise variance after sum aggregation is
\begin{equation}
    \Var_i \Big[\sum_{b \in \gN_a} \vm_{(ba)i} \Big] = \sum_{b \in \gN_a} \Var_i[\vm_{(ba)i}] + \sum_{b \in \gN_a} \sum_{c \in \gN_a \setminus \{b\}} \Cov_i[\vm_{(ba)i}, \vm_{(ca)i}].
    \label{eq:agg_var}
\end{equation}
This variance depends on the number of neighbors in $\gN_a$. However, we consistently found that rescaling the output depending on $\gN_a$ has negative effects on the accuracy. The likely reason for this is that atomic interactions scale roughly linearly with neighborhood size. Moreover, the covariance in \cref{eq:agg_var} is not zero since all messages $\vm_{ba}$ are transformed using the same weight matrices. We therefore best estimate this variance empirically.

For a Hadamard product-based message passing filter (and analogously for a bilinear layer) we have
\begin{equation}
    \Var_i [F_{i} \vm_{i}] = \Cov_i[F_{i}^2, \vm_{i}^2] + (\Var_i[F_{i}] + \E_i[F_{i}]^2) (\Var_i[\vm_{i}] + \E_i[\vm_{i}]^2) - ( \Cov_i[F_{i}, \vm_{i}] + \E_i[F_{i}] \E_i[\vm_{i}] )^2.
\end{equation}
The main problem with this covariance is the non-zero quadratic covariance $\Cov_i[F_{i}^2, \vm_{i}^2]$. We again estimate this variance empirically based on a data sample.

\begin{figure}
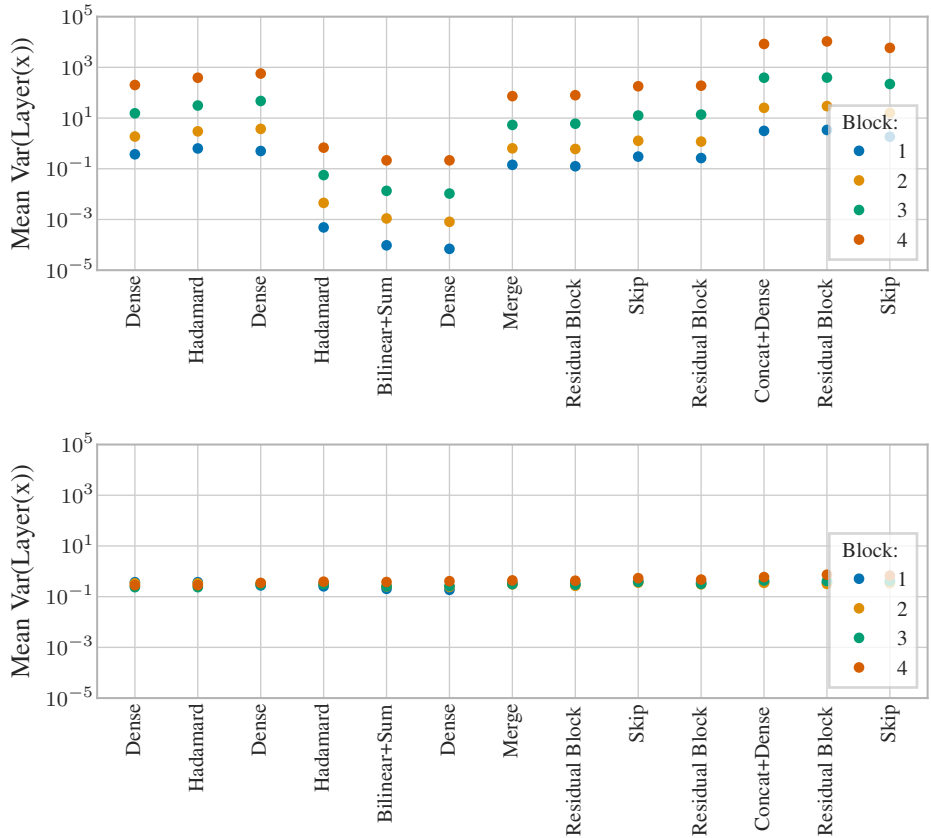

    \centering
    \input{figures/NotScaled.pgf}
    \input{figures/Scaled.pgf}
    \caption{Layer-wise activation variance. GemNet's variance varies strongly between layers and increases significantly after each block without scaling factors (top). Introducing scaling factors successfully stabilizes the variance (bottom).}
    \label{fig:var_scale}
\end{figure}

\FloatBarrier
\section{GemNet architecture} \label{app:gemnet}

We use 4 stacked interaction blocks and an embedding size of 128 throughout the model. For the basis functions we choose $N_{\text{SHBF}} = N_{\text{CHBF}} = 7$ and $N_{\text{SRBF}} = N_{\text{CRBF}} = N_{\text{RBF}} = 6$. For the weight tensor of the bilinear layer in the interaction block we use $N_{\text{bilinear,SBF}} = 32$ and $N_{\text{bilinear,CBF}} = 64$. We found that sharing the first weight matrix in \cref{eq:core_geom}, the down projection, resulted in the same validation loss but reduced the training time by up to \SI{15}{\percent}. The down projection size was chosen as 16 for the radial and circular basis and 32 for the spherical basis.

\begin{figure}[h!]
    \centering
    \resizebox{\textwidth}{!}{
    \input{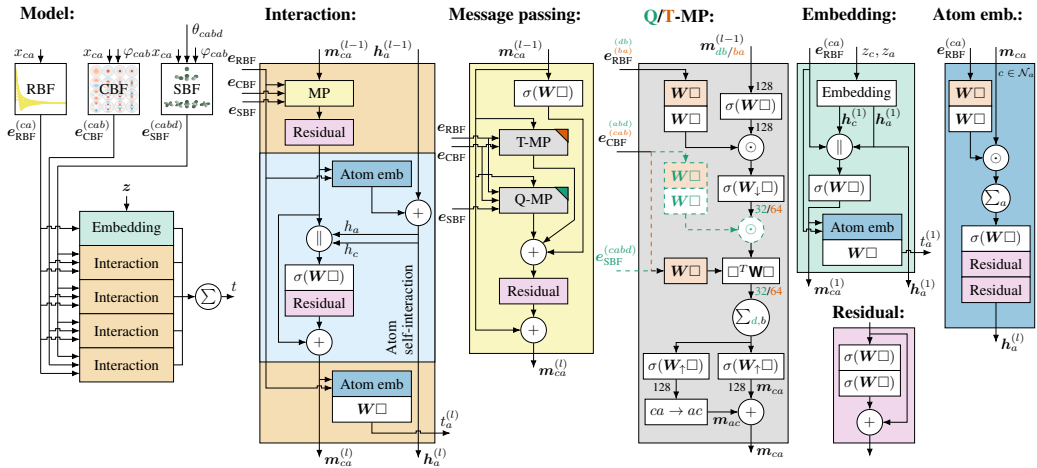}
    }
    \caption{The full GemNet architecture. $\square$ denotes the layer's input, $\|$ concatenation, $\sigma$ a non-linearity (we use SiLU in this work \citep{elfwing_sigmoid-weighted_2018}), and orange a layer with weights shared across interaction blocks. Differences between two-hop message passing (Q-MP) and one-hop message passing (T-MP) are denoted by dashed lines. Numbers next to connecting lines denote embedding sizes.}
    \label{fig:gemnet_full}
\end{figure}

\FloatBarrier
\section{Training and hyperparameters} \label{app:training}

\begin{table}[h!]
    \centering
    \caption{Model and training hyperparameters.}
    \begin{tabular}{lc@{\hspace{0.2cm}}c@{\hspace{0.2cm}}c}

Hyperparameters                 &       &        &        \\
\hline
Interaction cutoff $c_{\text{int}}$    & & \SI{10}{\angstrom}  & \\
Embedding cutoff $c_{\text{emb}}$      & & \SI{5}{\angstrom}   & \\
Learning rate                   & & \num{1e-3}          & \\
EMA decay                       & & \num{0.999}         & \\
Weight decay                    & & \num{2e-6}          & \\
Decay epochs                    & & \num{1200}          & \\
Decay rate                      & & \num{0.01}          & \\
Decay factor on plateau         & & \num{0.5}           & \\
Gradient clipping threshold     & & \num{10.0}          & \\
Envelope exponent               & & \num{5}             & \\
Force weighting factor $\rho$   & & \num{0.999}         & \\

\hline
                                &   MD17        &   MD17@CCSD(T)   &     Coll \\
\hline
Train set size                  & \num{1000}    & \num{950}     & \num{120000}\\
Val. set size                   & \num{1000}    & \num{50}      & \num{10000} \\
Max epochs                      & \num{2000}    & \num{2000}    & \num{400}   \\
Evaluation interval (epochs)    & \num{10}      & \num{10}      & \num{2}     \\
Decay on plateau patience (epochs) & \num{50}   & \num{50}      & \num{10}    \\
Decay on plateau cooldown (epochs) & \num{50}   & \num{50}      & \num{10}    \\
Warm-up epochs                  & \num{10}      & \num{10}      & \num{1}     \\
Batch size                      & \num{1}       & \num{1}       & \num{32}    \\
\end{tabular}

    \label{tab:hparams}
\end{table}

We subtract the mean energy from each molecule in MD17 to obtain a training target similar to atomization energy. We train on \si{\electronvolt} for energies and \si{\electronvolt\per\angstrom} for forces.
As a training objective we use the weighted loss function
\begin{equation}
    \mathcal{L}_\text{MD}(\mX, \vz) = (1-\rho) \left| f_\theta(\mX, \vz) - \hat{t}(\mX, \vz) \right| + \frac{\rho}{N} \sum_{i=1}^N \sqrt{\sum_{\alpha=1}^3 \left( -\frac{\partial f_\theta(\mX, \vz)}{\partial \vx_{i\alpha}} - \hat{F}_{i\alpha}(\mX, \vz) \right)^2},
\end{equation}
with force weighting factor $\rho = 0.999$.
We found the selection of the batch size to be of great influence on the model's performance for the MD17(@CCSD) dataset. Changing the batch size from 32 to 1 resulted in an approx. \SI{25}{\percent} lower validation MAE. The learning rate of $1\times 10^{-3}$ and the selection of the embedding cutoff $c_{emb} = \SI{5}{\angstrom}$ and interaction cutoff $c_{int} = \SI{10}{\angstrom}$ are rather important hyperparameters as well, see \cref{tab:cutoff}. We optimized the model using AMSGrad \citep{reddi_convergence_2018} with weight decay \citep{loshchilov_decoupled_2018} in combination with a linear learning rate warm-up, exponential decay and decay on plateau. However, we did not apply the weight decay for the initial atom embeddings, biases and frequencies (used in the radial basis). Without weight decay the force MAE was around \SI{3}{\percent} higher on COLL (not on OC20). Gradient clipping and early stopping on the validation loss were used as well.
In addition, we divided the gradients of weights that are shared across multiple blocks by the number of blocks the weights are shared for, which resulted in a small gain in accuracy. The model weights for validation and test were obtained using an exponential moving average (EMA) with decay rate \num{0.999}. The used hyperparameters can be found in \cref{tab:hparams}. The combined model on revised MD17 was trained with a batch size of 10.

We used a slightly adapted model for the OC20 dataset. It uses 128 Gaussian radial basis functions instead of spherical Bessel functions, which do not depend on the degree $l$ of the spherical harmonic. We furthermore used only three interaction blocks, an atom and edge embedding size of 512, an embedding cutoff of \SI{6}{\angstrom}, a learning rate of \num{5e-4}, no weight decay, only learning rate decay on plateau with a patience of \num{15000} steps and a factor of 0.8 (no warm-up or exponential decay), and a batch size of 2048.




\FloatBarrier
\section{Additional experimental results} \label{app:exp2}

\begin{table}[h!]
    \centering
    \begin{minipage}[t]{0.68\textwidth}
        \centering
        \caption{MAE for direct force predictions on MD17 in \si[per-mode=symbol]{\milli\electronvolt\per\angstrom}. The increased speed of direct force predictions comes at a significant cost of accuracy. Note that the direct models are still more accurate than many previous models.}
        \begin{tabular}{lS[table-format=1.1]@{\hspace{0.2cm}}S[table-format=1.1]@{\hspace{0.2cm}}S@{\hspace{0.2cm}}S}
                                    & \mcc{GemNet-Q}     &   \mcc{GemNet-T}     &   \mcc{GemNet-dQ}    &   \mcc{GemNet-dT}    \\
\hline
Aspirin                             &  9.4   &    9.5   &   17.8   &   18.0   \\
Benzene\citep{chmiela_machine_2017} &  6.3   &    6.3   &    8.5   &    8.0   \\
Benzene\citep{chmiela_towards_2018} &  1.5   &    1.4   &    2.5   &    2.3   \\
Ethanol                             &  3.8   &    3.7   &    6.4   &    6.8   \\
Malonaldehyde                       &  6.9   &    6.7   &   11.5   &   12.5   \\
Naphthalene                         &  2.2   &    2.4   &    5.2   &    5.9   \\
Salicylic acid                      &  5.4   &    5.5   &   12.9   &   13.2   \\
Toluene                             &  2.6   &    2.6   &    6.1   &    5.7   \\
Uracil                              &  4.5   &    4.2   &   11.7   &   10.9   \\
\end{tabular}

        \label{tab:direct}
    \end{minipage}
    \hfill
    \begin{minipage}[t]{0.3\textwidth}
        \centering
        \caption{Impact of the cutoff on force MAE on \textsc{COLL}. Results reported in \si[per-mode=symbol]{\milli\electronvolt\per\angstrom} after \num{500000} training steps.  Increasing the interaction cutoff to \SI{10}{\angstrom} slightly reduces the error. Decreasing the embedding cutoff to \SI{3}{\angstrom} significantly increases the error.}
        \begin{tabular}{S[table-format=2]@{\hspace{0.2cm}}S[table-format=2]@{\hspace{0.2cm}}S}
$c_{\text{emb}}/\si{\angstrom}$     &      $c_{\text{int}}/\si{\angstrom}$     &  \mcc{MAE} \\
\hline
5   & 10    &  27.0  \\
5   & 5     &  28.2  \\
3   & 10    &  33.4  \\
3   & 5     &  35.3  \\
\end{tabular}

        \label{tab:cutoff}
    \end{minipage}
\end{table}

\begin{table}[h!]
    \centering
    \begin{minipage}[t]{0.56\textwidth}
        \centering
        \caption{Force MAE for MD17 in \si[per-mode=symbol]{\milli\electronvolt\per\angstrom}. GemNet using \num{1000} training samples compared to SchNet using \num{50000} samples. GemNet outperforms SchNet on six out of eight molecules -- despite using 50x fewer samples.}
        \begin{tabular}{lSS[table-format=1.1]S[table-format=1.1]}
                                    & \mcc{SchNet 50k} & \mcc{GemNet-Q}           & \mcc{GemNet-T}            \\
\hline
Aspirin                             & 14.3 &   \bfseries  9.4   &    9.5   \\
Benzene\citep{chmiela_machine_2017} & 7.4  &   \bfseries  6.3   &   \bfseries  6.3   \\
Benzene\citep{chmiela_towards_2018} & \mcc{-} &    1.5   &   \bfseries  1.4   \\
Ethanol                             & \bfseries 2.2    &    3.8   &    3.7   \\
Malonaldehyde                       & \bfseries 3.5    &    6.9   &    6.7   \\
Naphthalene                         & 4.8  &   \bfseries  2.2   &    2.4   \\
Salicylic acid                      & 8.2  &   \bfseries  5.4   &    5.5   \\
Toluene                             & 3.9  &   \bfseries  2.6   &   \bfseries  2.6   \\
Uracil                              & 4.8  &    4.5   &   \bfseries  4.2
\end{tabular}

        \label{tab:md17_50k}
    \end{minipage}
    \hfill
    \begin{minipage}[t]{0.42\textwidth}
        \centering
        \caption{Effect of adding our independent improvements to DimeNet$^{++}$ on force MAE for COLL in \si[per-mode=symbol]{\milli\electronvolt\per\angstrom}. In this experiment we increased the basis embedding size of DimeNet$^{++}$ from 8 to 16 to eliminate this bottleneck. All improvements have a significant effect.}
        \begin{tabular}{lS}
    Model         &    \mcc{Forces} \\
    \hline
    DimeNet$^{++}$                                    &  41.1  \\
    with symmetric message passing                    &  37.5  \\
    with bilinear layer                               &  38.6  \\
    with scaling factors                              &  40.0  \\
\end{tabular}

        \label{tab:ablation_dimenet}
    \end{minipage}
\end{table}

\begin{table}[!ht]
    \centering
    \caption{Force MAE for the revised MD17 dataset \citep{christensen_role_2020} in \si[per-mode=symbol]{\milli\electronvolt\per\angstrom}. On average, GemNet outperforms FCHL19 by \SI{52}{\percent} and even UNiTE by \SI{5}{\percent}, which is a $\Delta$-ML approach based on quantum mechanical features \citep{qiao_unite_2021}.}

\begin{tabular}{lSSSS[table-format=2.1]}
                &   \mcc{FCHL19}    & \mcc{UNiTE}   &   \mcc{GemNet-Q} &   \mcc{GemNet-T} \\
\hline
Aspirin         &   20.9    & \bfseries 7.8    &    9.7    &    9.5    \\
Benzene         &    2.6    &  0.7              &    0.7    &   \bfseries 0.5    \\
Ethanol         &    6.2    &  4.2              &    \bfseries 3.6    &   \bfseries 3.6    \\
Malonaldehyde   &   10.3    &  7.1              &    6.7    &   \bfseries 6.6    \\
Naphthalene     &    6.5    &  2.4      &   \bfseries 1.9  &    2.1    \\
Salicylic acid  &    9.5    & \bfseries 4.1    &    5.3    &    5.5    \\
Toluene         &    8.8    &  2.9              &    2.3    &   \bfseries 2.2    \\
Uracil          &    4.2    & \bfseries 3.8    &    4.1    &   \bfseries 3.8    \\
\end{tabular}

    \label{tab:rmd17}
\end{table}

\begin{table}[!ht]
    \setlength{\tabcolsep}{0.1cm}
    \centering
    \caption{Force MAE of different models (number of parameters in parentheses) for the MD17 dataset in \si[per-mode=symbol]{\milli\electronvolt\per\angstrom}. GemNet performs worse with an embedding size of 64, but still substantially better than previous models with more parameters.}
    \begin{tabular}{lSSSS[table-format=1.1]}
& \mcc{PaiNN (600k)} & \mcc{DimeNet (1.9M)} & \mcc{GemNet-T 64 (490k)} & \mcc{GemNet-T (1.9M)} \\
\hline
Aspirin                             & 14.7    & 21.6    & 11.2          & \bfseries 9.5           \\
Benzene\citep{chmiela_machine_2017} & \mcc{-} & 8.1     & \mcc{-}       & \bfseries 6.3 \\
Benzene\citep{chmiela_towards_2018} & \mcc{-} & \mcc{-} & \bfseries 1.1 & 1.4           \\
Ethanol                             & 9.7     & 10.0    & 5.1           & \bfseries 3.7 \\
Malonaldehyde                       & 14.9    & 16.6    & 7.8           & \bfseries 6.7 \\
Naphthalene                         & 3.3     & 9.3     & 3.3           & \bfseries 2.4           \\
Salicylic acid                      & 8.5     & 16.2    & 6.9           & \bfseries 5.5           \\
Toluene                             & 4.1     & 9.4     & 3.3           & \bfseries 2.6 \\
Uracil                              & 6.0     & 13.1    & 5.3           & \bfseries 4.2
\end{tabular}

    \label{tab:md17_small}
\end{table}

\begin{table}[!ht]
    \centering
    \caption{Force MAE of GemNet on the revised MD17 dataset \citep{christensen_role_2020} in \si[per-mode=symbol]{\milli\electronvolt\per\angstrom} when using individual models for each molecule (``Individual'') versus a single model for all molecules (``Combined''). The combined setting is harder to learn, leading to a higher error in most cases. GemNet-Q performs better than GemNet-T in this setting.}
    \begin{tabular}{lSSSSS}
    & \multicolumn{2}{c}{GemNet-Q} & & \multicolumn{2}{c}{GemNet-T} \\
    \cline{2-3} \cline{5-6}
    & \mcc{Individual \rule{0pt}{0.9em}}    & \mcc{Combined}   & & \mcc{Individual}   & \mcc{Combined}   \\
\hline
Aspirin \rule{0pt}{0.9em}        &  9.7          &   10.0      & &    9.5       &    9.9    \\
Benzene        &  0.7          &    0.5      & &    0.5       &    0.6    \\
Ethanol        &  3.6          &    4.4      & &    3.6       &    4.9    \\
Malonaldehyde  &  6.7          &    7.7      & &    6.6       &    8.3    \\
Naphthalene    &  1.9          &    1.9      & &    2.1       &    2.2    \\
Salicylic acid &  5.3          &    4.6      & &    5.5       &    5.0    \\
Toluene        &  2.3          &    2.2      & &    2.2       &    2.5    \\
Uracil         &  4.1          &    4.1      & &    3.8       &    4.3    \\
\end{tabular}


    \label{tab:md17_combined}
\end{table}

\FloatBarrier
\section{Computation time} \label{app:comp_time}

\begin{table}[h!]
    \centering
    \sisetup{table-format=1.3}
    \caption{Runtime per batch of Toluene molecules on an Nvidia GeForce GTX 1080Ti in seconds. GemNet-T is comparably fast to previous methods. Note that NequIP requires roughly 10x more training epochs than GemNet for convergence \citep{batzner_se3-equivariant_2021}. Using direct force predictions and only one-hop message passing significantly accelerates training and inference (GemNet-dT). Efficient aggregation allows for the usage of a bilinear layer instead of a Hadamard product at no additional cost (GemNet-Q vs.\ Hadamard-Eff) and enables training with higher batch sizes (Hadamard-Eff vs.\ Hadamard-NonEff). Note that our implementation does not focus on runtime and can likely be significantly optimized.}
    \begin{tabular}{lSSSSS}
    & \multicolumn{2}{c}{batch size 32} & & \multicolumn{2}{c}{batch size 4} \\ \cline{2-3} \cline{5-6}
    & \mcc{Training \rule{0pt}{0.9em}}    & \mcc{Inference}   & & \mcc{Training}   & \mcc{Inference}   \\ \hline
DimeNet$^{++}$            & 0.357       & 0.065       & & 0.283      & 0.031       \\
NequIP (l=1)              & 0.066       & 0.042       & & 0.070      & 0.044       \\
NequIP (l=3, reflections) & 0.336       & 0.206       & & 0.327      & 0.197       \\
\hline
GemNet-Q                  & 1.067       & 0.376       & & 0.628      & 0.099       \\
GemNet-T                  & 0.397       & 0.088       & & 0.299      & 0.038       \\
GemNet-dQ                 & 0.369       & 0.264       & & 0.106      & 0.052       \\
GemNet-dT                 & 0.134       & 0.067       & & 0.065      & 0.020       \\
\hline
Hadamard-Eff              & 1.077       & 0.392       & & 0.632      & 0.103       \\
Hadamard-NonEff           & \mcc{OOM}   & 0.378       & & 0.633      & 0.103
\end{tabular}

    \label{tab:runtimes}
\end{table}

The models were trained primarily using Nvidia GeForce GTX 1080Ti GPUs. For MD17 and MD17@CCSD training the direct force prediction variants took less than two days, GemNet-Q and GemNet-T took around 6 days per molecule but with very little progress after the 100 hour mark. However, thanks to the memory efficient implementation and the low batch size used, several models were trained in parallel on a single GPU. On the \textsc{COLL} dataset training the direct force prediction variants took around 24 hours each. GemNet-T trained for 60 hours, while GemNet-Q took 6 days. However, after 60 hours GemNet-Q is already within \SI{5}{\percent} of its final validation error and outperforms GemNet-T by a large margin. Note that the training time reduces dramatically when using a larger batch size, at the cost of a slightly higher MAE on MD17.

\end{document}